%% file: main.tex
\documentclass[letterpaper]{article}
\usepackage{aaai}
\usepackage{times}
\usepackage{helvet}
\usepackage{courier}
\usepackage{graphicx, subfigure, comment, array,booktabs,arydshln,xcolor}
\usepackage{textcomp}
\usepackage{multirow}
\newcommand{\hide}[1]{}

\begin{document}
%


\title{Psycho-Demographic Analysis of the Facebook Rainbow Campaign}

%
%
%
%
%

\author{Submission 99}
\author{Yilun Wang$^1$, Himabindu Lakkaraju$^2$, Michal Kosinski$^3$, Jure Leskovec$^4$\\
$^{1,2,4}$Computer Science Department, Stanford University\\
$^3$Graduate School of Business, Stanford University\\
$^{1,2,4}$\{yilunw, himalv, jure\}@cs.stanford.edu, $^3$michalk@stanford.edu
}

\maketitle
  \begin{abstract}
Over the past decade, online social media has had a tremendous impact on the way people engage in social activism. For instance, about 26M Facebook users expressed their support in upholding the cause of marriage equality by overlaying their profile pictures with rainbow-colored filters. Similarly, hundreds of thousands of users changed their profile pictures to a black dot condemning incidents of sexual violence in India. This act of demonstrating support for social causes by changing online profile pictures is being referred to as \emph{pictivism}. \footnote{\tiny{http://www.impatientoptimists.org/Posts/2013/03/Pictivism-Did-You-Change-Your-Facebook-Pic-to-Red-This-Week}} In this paper, we analyze the psycho-demographic profiles, social networking behavior, and personal interests of users who participated in the Facebook Rainbow campaign. Our study is based on a sample of about 800K detailed profiles of Facebook users combining questionnaire-based psychological scores with Facebook profile data. Our analysis provides detailed insights into psycho-demographic profiles of the campaign participants. We found that personality traits such as openness and neuroticism are both positively associated with the likelihood of supporting the campaign, while conscientiousness exhibited a negative correlation. We also observed that females, religious disbelievers, democrats and adults in the age group of 20 to 30 years are more likely to be a part of the campaign. Our research further confirms the findings of several previous studies which suggest that a user is more likely to participate in an online campaign if a large fraction of his/her friends are already doing so. We also developed machine learning models for predicting campaign participation. Users' personal interests, approximated by Facebook user like activity, turned out to be the best indicator of campaign participation. Our results demonstrated that a predictive model which leverages the aforementioned features accurately identifies campaign participants (AUC=0.76).

\end{abstract} 
\section{Introduction}
\label{sec:intro}
\input{010introduction.tex}

\section{Related Work}
\label{sec:relwork}
\input{080relwork.tex}

\section{Data}
\label{sec:dataset}
\input{020dataset.tex}

\section{Empirical Analysis}
\label{sec:emp_ana}
\input{030demog.tex}

\subsection{Personality}
\label{sec:personality}
\input{040personality.tex}

\subsection{Social Networks}
\label{sec:social}
\input{050social.tex}

\subsection{Interests and Preferences}
\label{sec:like}
\input{060like.tex}

\section{Predicting Campaign Participation}
\label{sec:prediction}
\input{070prediction.tex}

\section{Conclusions \& Discussion}
\label{sec:conclusion}
\input{090conclusion.tex}


%
\bibliographystyle{aaai}
\small
\bibliography{sigproc}
\normalsize
%
%

\end{document}

%% file: 010introduction.tex
The advent of online social media has fueled a change in the practice of social activism. Ease and speed of communication, massive reach, and the interactive nature of social media proved to be immensely helpful in expediting social activism~\cite{obar2012advocacy}. People can now express their opinions about social and political issues with just a mouse click, thanks to features such as "like", "share", and "tweet" on Facebook and Twitter~\cite{conover2013geospatial,tilly2013social,gonzalez2011dynamics}. 

\begin{figure}
    \centering
    \includegraphics[width=0.99\linewidth]{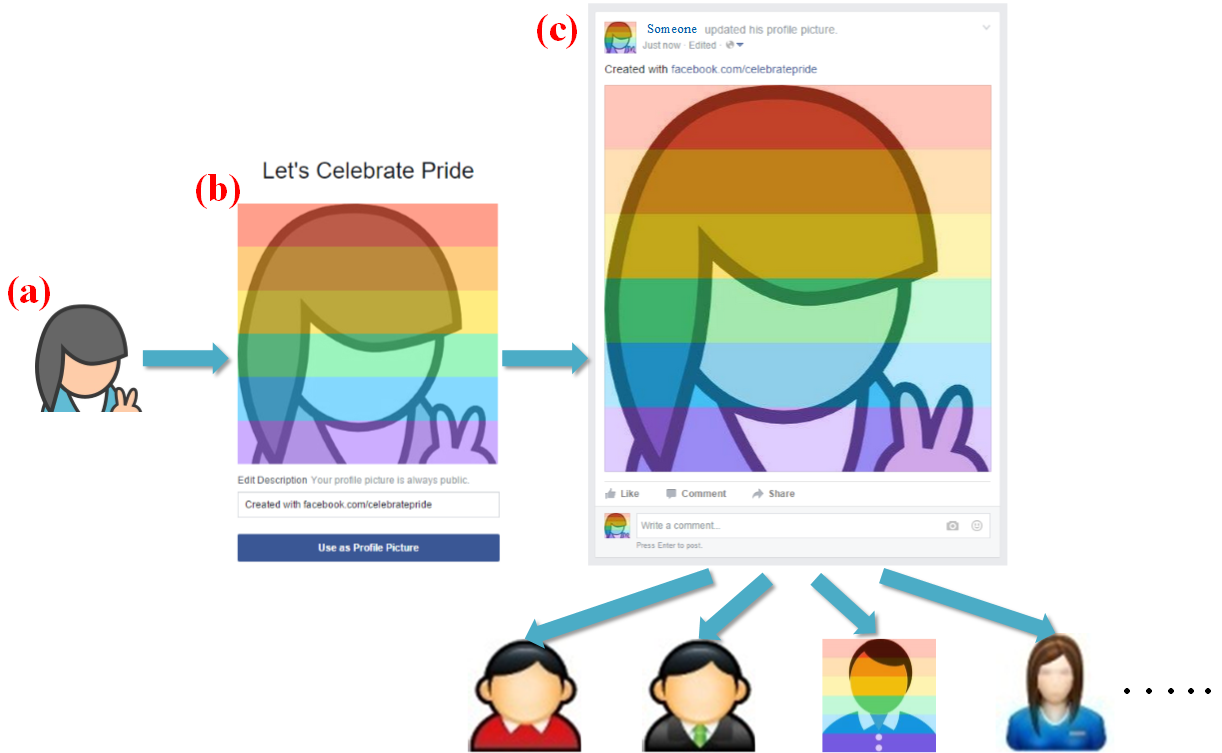}
    \caption{Facebook rolled out a tool enabling users to join the Rainbow campaign and facilitating its viral spread. With a single click, users could overlay their profile pictures (a) with a rainbow filter (b) post an update on their Facebook wall which contains a link to the profile picture changing tool.}
    \label{fig:fb}
    \vspace{-0.1in}
\end{figure}

Pictivism, a portmanteau of the words \textit{picture} and \textit{activism}, represents a new form of activism where online users show support for a cause by overlaying their profile pictures with particular filters, ribbons, or badges. Though the effectiveness of such campaigns in bringing about social changes is debatable, they have gained enormous attention in the recent past. This is mainly due to the fact that such campaigns require minimal personal effort from participants~\cite{carr2012hashtag}.
The Facebook Rainbow campaign is one of the most massive and most recent instances of pictivism: 26 million users reportedly applied a rainbow-colored filter to their profile pictures after the U.S. Supreme Court made same-sex marriage legal nationwide in June 2015.
Thanks to the profile picture changing tool provided by Facebook, several users celebrated the landmark Supreme Court decision by slapping rainbow flags on their profile pictures with just one click. In addition, friends of these users could view all the profile picture updates in their news feeds which in turn attracted more users to join the campaign (See Figure~\ref{fig:fb}).

Several previous studies focused on understanding how demographic traits, such as age, gender, and location \cite{herek1988heterosexuals,pearl2007development}, political views \cite{sherkat2011religion}, and religious views \cite{olson2006religion,sherkat2011religion} influence a person's decision to support same-sex marriage or LGBT rights more broadly. These studies, however, were based on small and relatively biased samples. They also relied on self-reported participation in the cause, which might not be an accurate reflection of actual behavior~\cite{gonyea2005self}. Furthermore, no prior research has attempted to understand the psychological traits of people who support same-sex marriage. Similarly, there have been no studies which explored the psychological characteristics of users who participate in online campaigns.
\\

\noindent \textbf{Present work:} In this paper, we analyze the Facebook Rainbow campaign, an unanticipated, massive, natural experiment that followed the U.S. Supreme Court's decision to legalize same-sex marriage. The public nature of the Facebook Rainbow campaign 
allowed us to directly observe users' participation, rather than rely on self-reported measures. We utilize a large sample of about 800K U.S. users obtained from the myPersonality database. Studies have demonstrated that the user sample captured by myPersonality database is indeed representative of Facebook population~\cite{Kosinski_2015}. To detect users who participated in the campaign, we scraped and analyzed their Facebook profile pictures to check for the presence of the rainbow filter.\footnote{myPersonality users provided an opt-in consent allowing the usage of their Facebook profile data for research purposes. Profile pictures were used solely to ascertain user participation in the Rainbow campaign.} 
Since our analysis is carried out on a large-scale, real-world dataset of U.S. Facebook users, the insights that we obtain are likely to generalize across larger population.

Our results provide detailed insights into psycho-demographic profiles of participants in the Facebook Rainbow campaign. In particular, we analyze how demographic attributes, personality, happiness, intelligence, personal interests, and social network structure influence the campaign participation. Additionally, we develop machine learning models to predict campaign participation based on these factors and in turn utilize these models to explore the predictive power of various combinations of the aforementioned features. To the best of our knowledge, this work is the first attempt to analyze the relationship between such a broad range of psycho-demographic traits and participation in a massive online campaign.

%% file: 080relwork.tex
We begin this section by summarizing all the sociological studies which document public opinions on same-sex marriage and also explain how socio-demographic factors are known to influence peoples' opinions regarding the same. Next, we discuss various case studies of online activism and briefly mention prior work which characterized user participation in online campaigns. 

\paragraph{Same-sex marriage} Over the past couple of decades, same-sex marriage has been a very important topic of study in sociology and behavioral psychology. While some of the prior research focused on documenting public opinion trends on same-sex marriage and more broadly, LGBT rights~\cite{hicks2006public,yang1997trends}, other studies attempted to identify the underlying factors which determine attitudes towards this controversial subject~\cite{brumbaugh2008attitudes,campbell2008religion,gaines2010morality,herek1988heterosexuals,herek2002gender}. More specifically, prior research concluded that youngsters~\cite{pearl2007development}, females~\cite{herek1988heterosexuals,herek2002gender}, and democrats~\cite{olson2006religion,sherkat2011religion} were more likely to be accepting towards the LGBT population and consequently were supporters of the same-sex marriage.   

Several of the aforementioned studies, however, were carried out on a relatively small sample of users and some of them even relied only on data from very few demographic groups. Furthermore, almost of all these studies employed traditional methods for collecting opinions such as surveys and questionnaires. For instance, \citeauthor{gaines2010morality} analyzed data obtained from a national survey with 620 participants. \citeauthor{brumbaugh2008attitudes} utilized data collected from surveys conducted in three states of the United States. This implies that the obtained insights were representative of a specific and biased user sample. Also, answers to survey questions which determine respondents' willingness to support a cause are often subject to unintended and/or willful misrepresentation
~\cite{rust2014modern}. 
\vspace{-0.15in}
\paragraph{Online activism} 
Due to the thriving presence of a variety of online portals, social activism has found expression in various forms such as updating profile pictures to demonstrate support for a cause (i.e. pictivism), creating pages on Facebook and encouraging users to \emph{like} them thus increasing awareness about a social issue, creating petitions on portals such as Change.org and asking users to sign them indicating their support, tweeting about a cause etc.

Several studies have focused on understanding the dynamics of specific online movements. For instance, \citeauthor{lewis2014structure} presented details about the online recruitment of activists and the associated diffusion process pertaining to the \emph{Save Darfur Cause}, which intended to raise awareness about various atrocities occurring in a region called \emph{Dafur} in Sudan. Similarly, detailed studies were published on several online campaigns such as Occupy Wall Street~\cite{conover2013geospatial}, Arab Spring~\cite{tilly2013social}, Spanish 15-M movement~\cite{gonzalez2011dynamics}, the Moldavian Twitter Revolution~\cite{mungiu2009moldova} and the Guatemalan justice movement~\cite{harlow2012social}. Few studies focused on characterizing the geographical aspects of online movements~\cite{conover2013geospatial}, while others analyzed the association between demographic attributes and campaign participation, and the associated diffusion dynamics~\cite{adamic2015diffusion,lewis2014structure}. 
Huang et. al.~\cite{huang2015activists} analyzed petition data from Change.org and presented an in-depth characterization of successful online campaigns and activists. 

Though most of the aforementioned research utilized large volumes of online data, these analyses did not involve the study of the influence of psychological traits, religious views and political views on campaign participation. Furthermore, very few studies have dealt with pictivism, the most recently discovered form of activism.

%% file: 020dataset.tex
\begin{table}
\centering
\small
\caption{Descriptive statistics of variables used.}
\begin{tabular}{rll} 
\hline
Variable& Users & Details \\
\hline 
\\
Total \# of users & 799,091 & 6.96\% joined the campaign\\ \\
Gender & 785,176 & 65.42\% female\\ 
Age & 542,182 & Range: 18-60 years old\\
&& Median age: 26 years old\\
Political views & 249,703 & Liberal (41.84\%)\\
&&Conservative (24.38\%)\\
Religion & 372,168 & Christian (69.82\%) \\
&&Non-believer (12.91\%)\\ 
Location & 431,626 & Range: U.S. states \\ \\
Personality (FFM) & 799,091 & Openness\\
&&Conscientiousness\\
&&Extroversion\\
&&Agreeableness\\
&&Neuroticism \\ 
Happiness & 24,965 & Average: 0.0\\
Intelligence  &2,357& Average: 112.38\\ \\
Social Network Ties & 178,660 & Ties: 210,118\\ \\
Facebook Likes & 50,520 & Unique Likes: 35,284 \\ 
&&User-like diads: 14,227,568\\
&&Median Likes per user: 143\\ 
&&Median users per Like: 133\\  \\ 
\hline
\end{tabular}
\label{tab:desc_stat}
\end{table}
\normalsize
This study is based on a sample of 799,091 U.S. Facebook users obtained from the myPersonality database.\footnote{http://www.mypersonality.org/} myPersonality was a popular Facebook application that provided feedback on users' psychological profiles in the form of scores computed via a wide range of psychometric tests. About 3.5 million users used this application and provided an opt-in consent indicating that the data from their psychological and Facebook profiles can be used for research purposes. In this study, we focused exclusively on the U.S. users in the myPersonality database since this population was impacted by the U.S. Supreme Court's judgement favoring legalization of same-sex marriage. Previous research on myPersonality database showed that its users belonged to a wide range of demographic groups, backgrounds and cultures, thus providing us with a reasonably representative sample of Facebook population~\cite{Kosinski2013ml,Kosinski_2015}. With users' consent, their profile information, social network ties and like activity on Facebook, and several other attributes such as personality scores, happiness levels, intelligence quotient, were recorded during the time period between 2009 and 2011. Profile pictures which were used to detect rainbow filters were retrieved in July 2015. 
\vspace{-0.15in}
\paragraph{Facebook Profile Information} 
A wide variety of demographic attributes were recorded from users' Facebook profiles, including age, gender, location (U.S. state), political view, and religion. See Table~\ref{tab:desc_stat} for descriptive statistics of variables used in this study.
Our analysis was restricted to users in the age group of 18 to 60 years, which accounted for about 98.73\% of users who provided their age.

Religious views of most users can be categorized as: Christians (including a broad range of denominations) and non-believers (including ``atheists'' and ``agnostics''). For the sake of simplicity, we limited this variable to those two categories (accounting for 82.73\% of the users with non-missing values for religion) and discarded other, less popular, religious views. Similar prepossessing method was applied to political views where 66.22\% of users could be categorized into two groups: liberals (``liberal'', ``very liberal'', ``democrat'')  and conservatives (``conservative'', ``very conservative'', and ``republican'').

In our analysis, we made no assumptions about datapoints with missing values. Whenever we analyzed a particular attribute or combination of attributes in the data and studied their relationship with the participation in the Rainbow campaign, we ignored the datapoints that had a missing value associated with any of the corresponding attribute. \vspace{-0.15in}  
\paragraph{Personality, Happiness, and Intelligence}
Users' personality traits were measured in accordance with the Five Factor Model of Personality (FFM)~\cite{Costa1992neo}. Each of the 800K users was assigned scores on attributes such as openness, conscientiousness, extraversion, agreeableness, and neuroticism. These scores were measured using a standard questionnaire called 
International Personality Item Pool (IPIP)~\cite{goldberg2006international}. IPIP has been widely used in both traditional and online personality assessments, and is known to provide reliable estimates~\cite{buchanan2005implementing,chernyshenko2001fitting}. 

Happiness was measured using Satisfaction With Life Scale (SWLS) which is a short 5-item instrument widely used in research on subjective well-being~\cite{diener1985satisfaction}. This data was available for 24,965 users in our sample. 

Intelligence was measured using Raven's
Standard Progressive Matrices (SPM)~\cite{raven2000raven}, a multiple choice nonverbal intelligence test drawing on Spearman's theory of general ability. This test is a proven standard intelligence test used in both research and clinical settings, as well as in high-stake contexts such as in military personnel selection and court cases. 
This information was available for 2,357 users in our sample. 
\vspace{-0.15in}
\paragraph{Social Network Ties}
Social network ties were extracted from users' Facebook friendship networks. Note that due to privacy concerns, we did not study full egocentric networks but merely the connections between individuals in our sample. We observed 210,118 friendship links in total.
The available network information serves as a useful way for approximating users' egocentric networks.   
\vspace{-0.15in} 
\paragraph{Facebook Likes}
Facebook users leverage the \textit{like} feature to express their preferences, interests, or support. Users can like a variety of entities including but not limited to websites, institutions, places, products, movies, books, and artists. From here on we refer to all such entities which can be liked by users on Facebook as \emph{Likes} (with capitalization).
We retained only those users who had at least 50 Likes (50,520 users) and only those Likes which were associated with at least 50 users (35,284 Likes) to avoid noise in the data. 
\vspace{-0.15in} 
\paragraph{Detecting Rainbow Filters}
Facebook Rainbow campaign participants were identified based on their profile pictures using a simple rule-based computer vision algorithm which analyzed the color distribution in various regions of the picture. In order to evaluate this algorithm, we manually analyzed a random sample of 19,619 profile pictures and identified pictures with rainbow filter. Comparing the computer vision algorithm with human baseline, the algorithm produced zero false positives and just two false negatives, indicating a classification precision of 100\% and a recall of 99.9\%. 

Due to the large volume of users in myPersonality database, it took us four days (between the 1st and 4th of July 2015) to record all their profile pictures using Facebook Graph API.\footnote{The campaign began on June 26th.} 
As Facebook users applied (and removed) the rainbow filter at different times, not all of the campaign participants were detected in this study. 
Overall, we found that $6.96\%$ of the users in our dataset applied a rainbow filter to their profile pictures.

%% file: 030demog.tex
\begin{figure}
\centering
\includegraphics[width=0.99\linewidth]{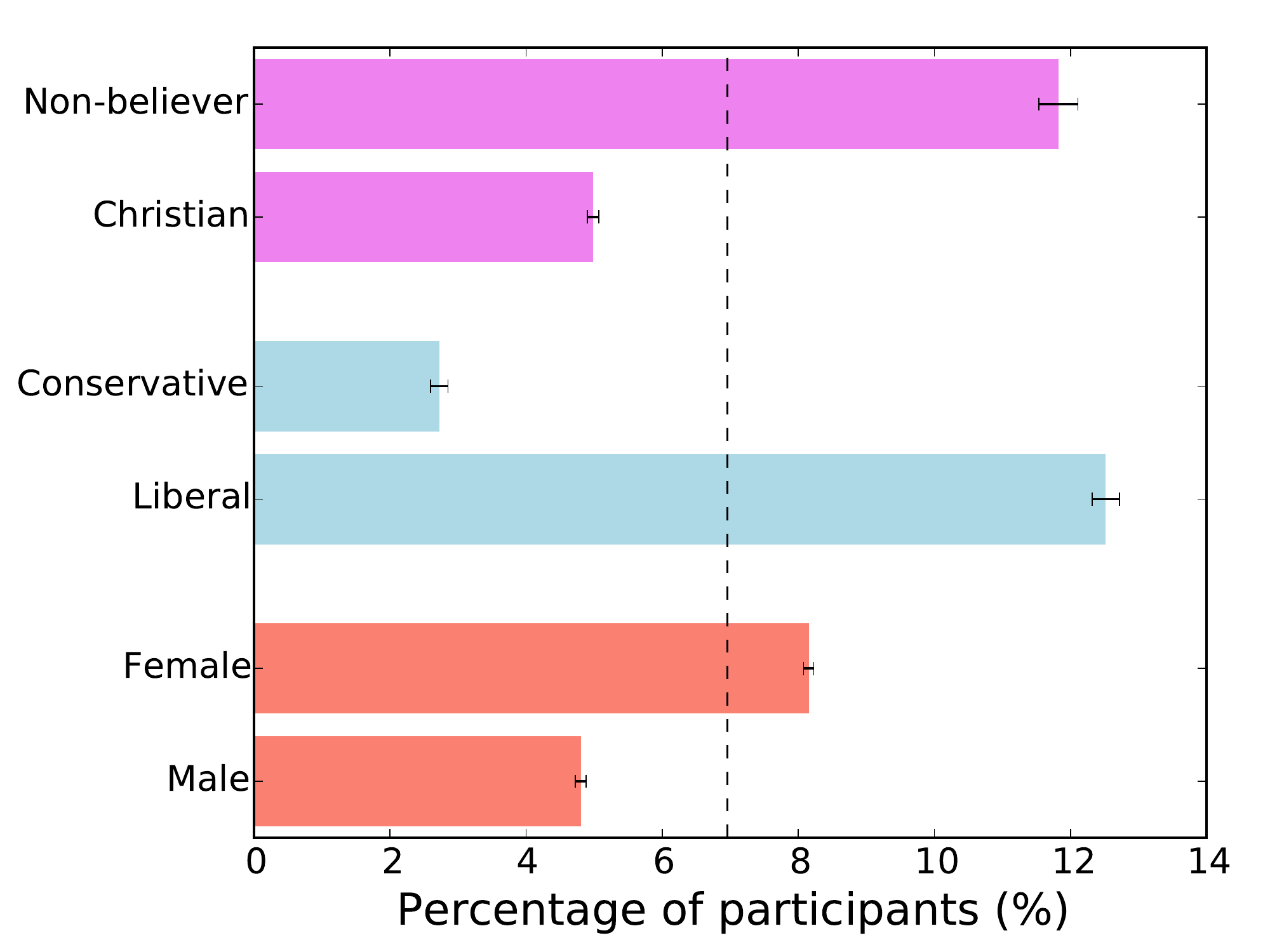}
\caption{Percentage of Rainbow campaign participants across religion, political inclination, and gender. Error bars represent the 95\% confidence interval and the dashed line represents the overall percentage of rainbow-colored profile pictures in our sample (6.96\%).}
\label{fig:rb_all}
\end{figure}

In this section, we investigate the relationship between campaign participation and several user features such as demographic attributes, religion, political views, personality traits, happiness, intelligence, social network ties, and personal interests.

\subsection{Age, gender, religion, and political views}
\label{sec:demog}
We start with an analysis of the campaign participants' gender, age, religion, and political views. 
Previous studies showed that women, liberals, and non-believers are generally more supportive of same-sex marriages~\cite{pew2015changing,sherkat2011religion,adamic2015diffusion}. Our results are consistent with those findings. Figure \ref{fig:rb_all} shows that women (8.15\%), liberals (12.52\%), and non-believers (11.82\%) are more likely to participate in the campaign compared to males (4.80\%), conservatives (2.72\%), and Christians (4.99\%).  

\begin{figure}
\centering
\includegraphics[width=0.99\linewidth]{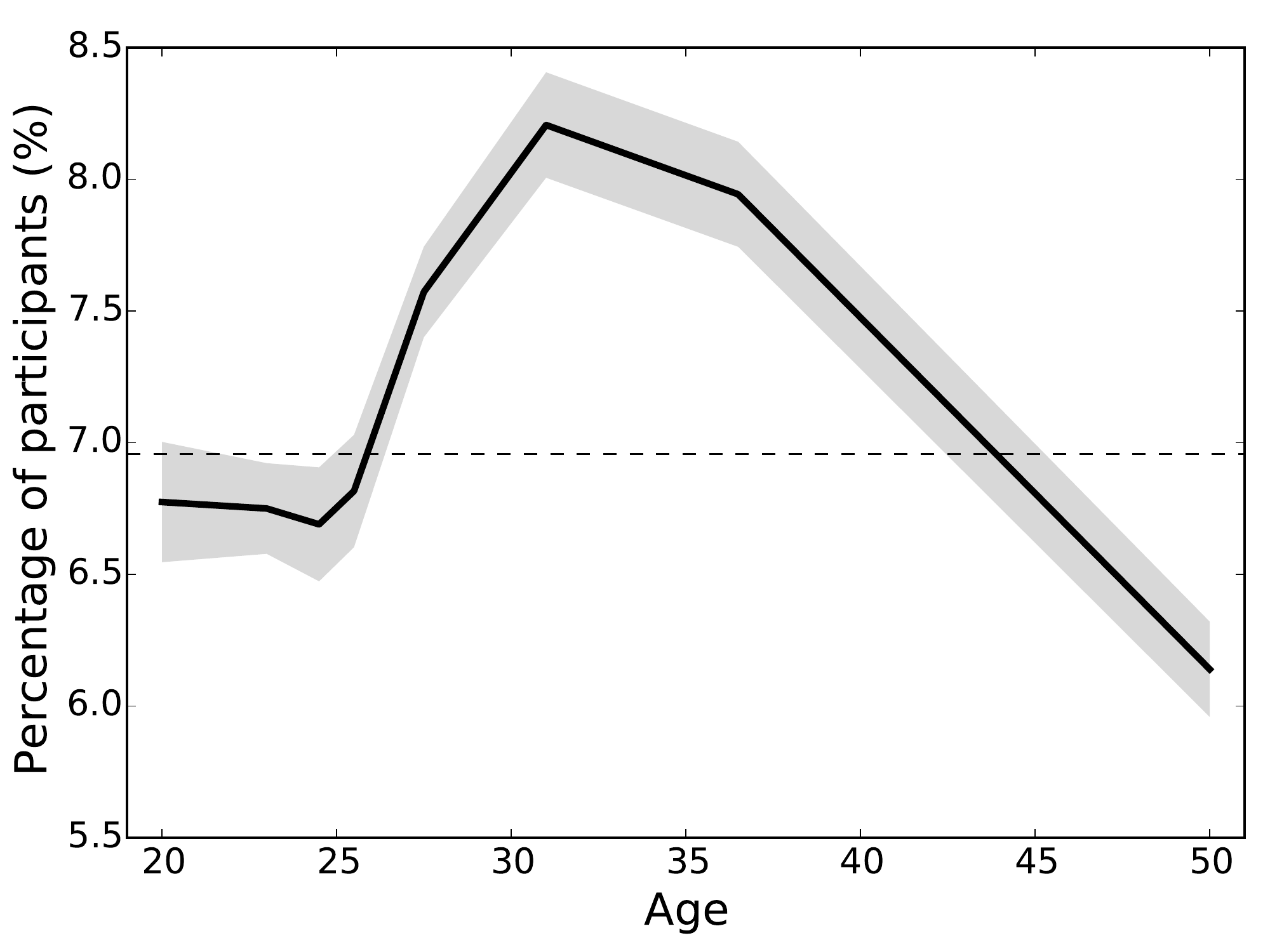}
\vspace{-0.2in}
\caption{Relationship between age and the Rainbow campaign participation. Dashed line represents the overall percentage of profile pictures with a rainbow filter in our dataset (6.96\%). Grey ribbon represents the 95\% confidence interval.}
\label{fig:rb_age}
\vspace{-0.2in}
\end{figure}

The relationship between age and campaign participation is presented in Figure \ref{fig:rb_age}. For the purpose of this analysis, we binned the values taken by age variable into eight quantiles. While prior research employing surveys or questionnaires showed that the younger population (adolescents and people in their early 20s) is more supportive of homosexuality~\cite{andersen2008cohort} and same-sex marriage~\cite{pew2015changing}, our results indicate that adults ranging between 30 to 40 years of age are more likely (8.21\%) to join the campaign compared to both adolescents (6.77\%) and senior citizens (6.14\%). Surprisingly low levels of participation among adolescents could be, potentially, explained by the social pressures experienced by teenagers who support same-sex marriage~\cite{almeida2009emotional}.

\subsection{Location}
\label{sec:location}
\begin{figure*}[ht] 
    \centering
    \subfigure[]{\includegraphics[width=0.53\textwidth]{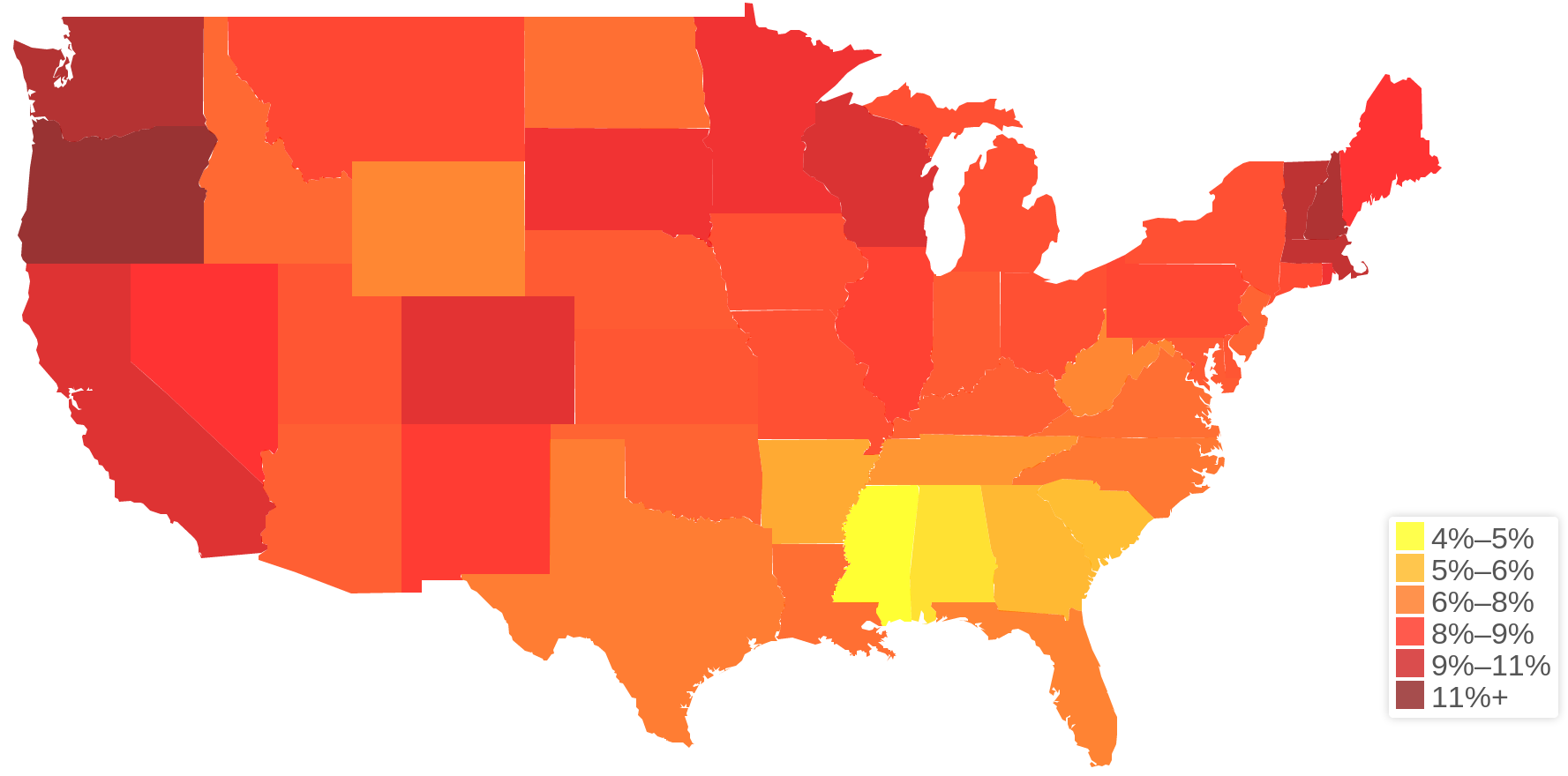}}
    \subfigure[]{\includegraphics[width=0.45\textwidth]{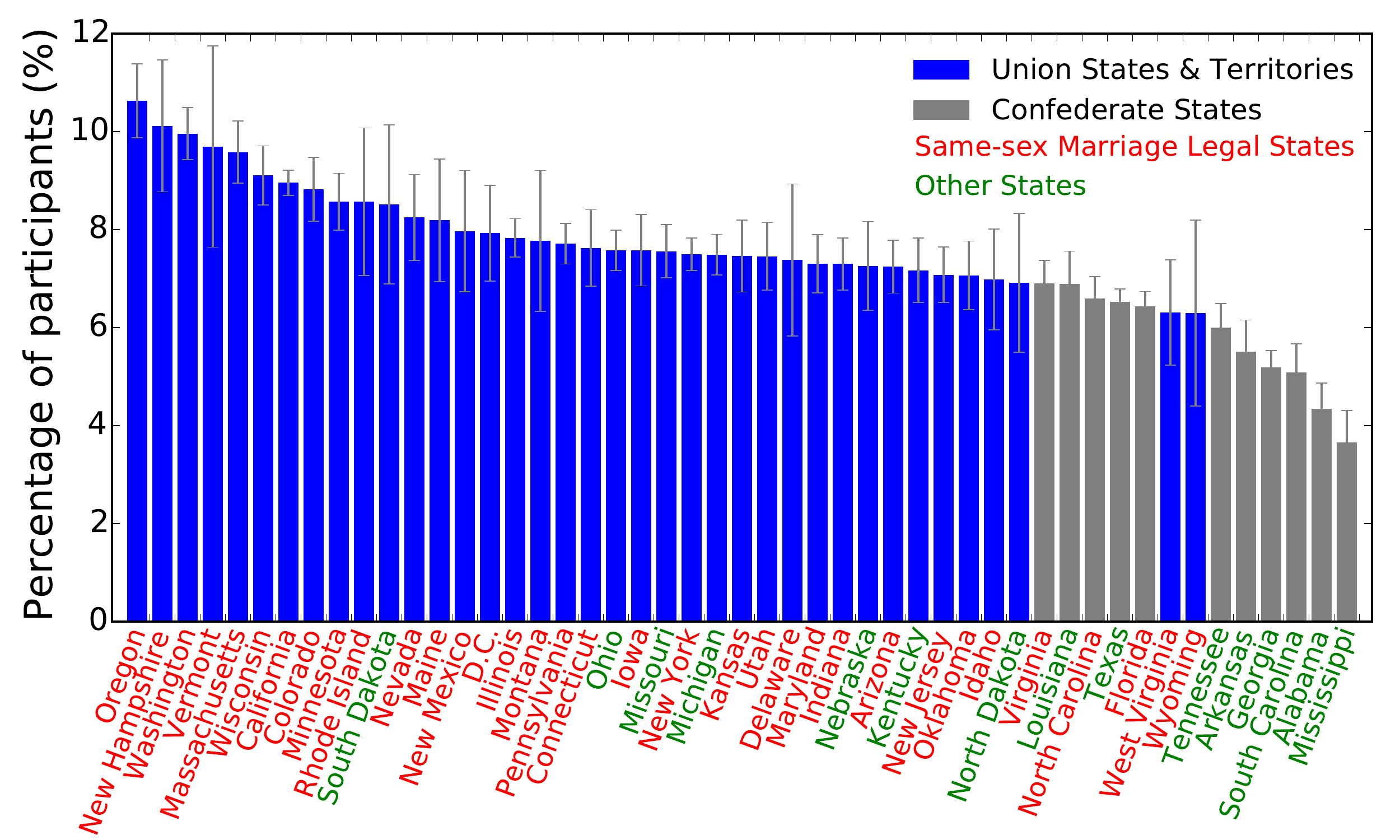}}
    \caption{Geographical distribution of campaign participation across the U.S. states represented as (a) a map and (b) a bar-chart. Blue bars represent Union states and grey bars represent Confederate states. States where the same-sex marriage was legal prior to the ruling are represented by red labels (on the x-axis), while other states are represented by green labels.}
    \label{fig:rb_state}
    \vspace{0.05in}
\end{figure*}

Figure~\ref{fig:rb_state}a highlights the geographical distribution of campaign participation. This map reveals a clear divide between the U.S. states, which closely mimics the historical divide between the Union and Confederate States. In fact, if the states are ordered based on the participation (See Figure~\ref{fig:rb_state}b and blue/grey bars), only two Union states (Wyoming and West Virginia) are located on the right, ``Confederate'', side of the plot and one of these states joined the Union only on June 20, 1863 (Virginia). In fact, it implies that campaign participation alone enables us to distinguish between Union and Confederate states with a very high accuracy (AUC$=0.98$).\footnote{AUC stands for Area Under Receiver Operating Characteristic Curve. The AUC is equivalent to the probability of correctly classifying two randomly selected states from each class.} 

We further explore the relationship between the campaign participation from a particular state and the legal status of same-sex marriage in that state prior to the U.S. Supreme Court verdict. States where same-sex marriage was legal prior to the ruling are marked with red text labels on the x-axis in Figure \ref{fig:rb_state}b, while other states are denoted using green text labels. It is clear that users from the states that have legalized same-sex marriage prior to the ruling were more supportive of the campaign. We found that campaign participation can accurately identify the states where same-sex marriage was legal before the ruling ({AUC}$=0.81$). 
\begin{figure}
\centering
\includegraphics[width=0.99\linewidth]{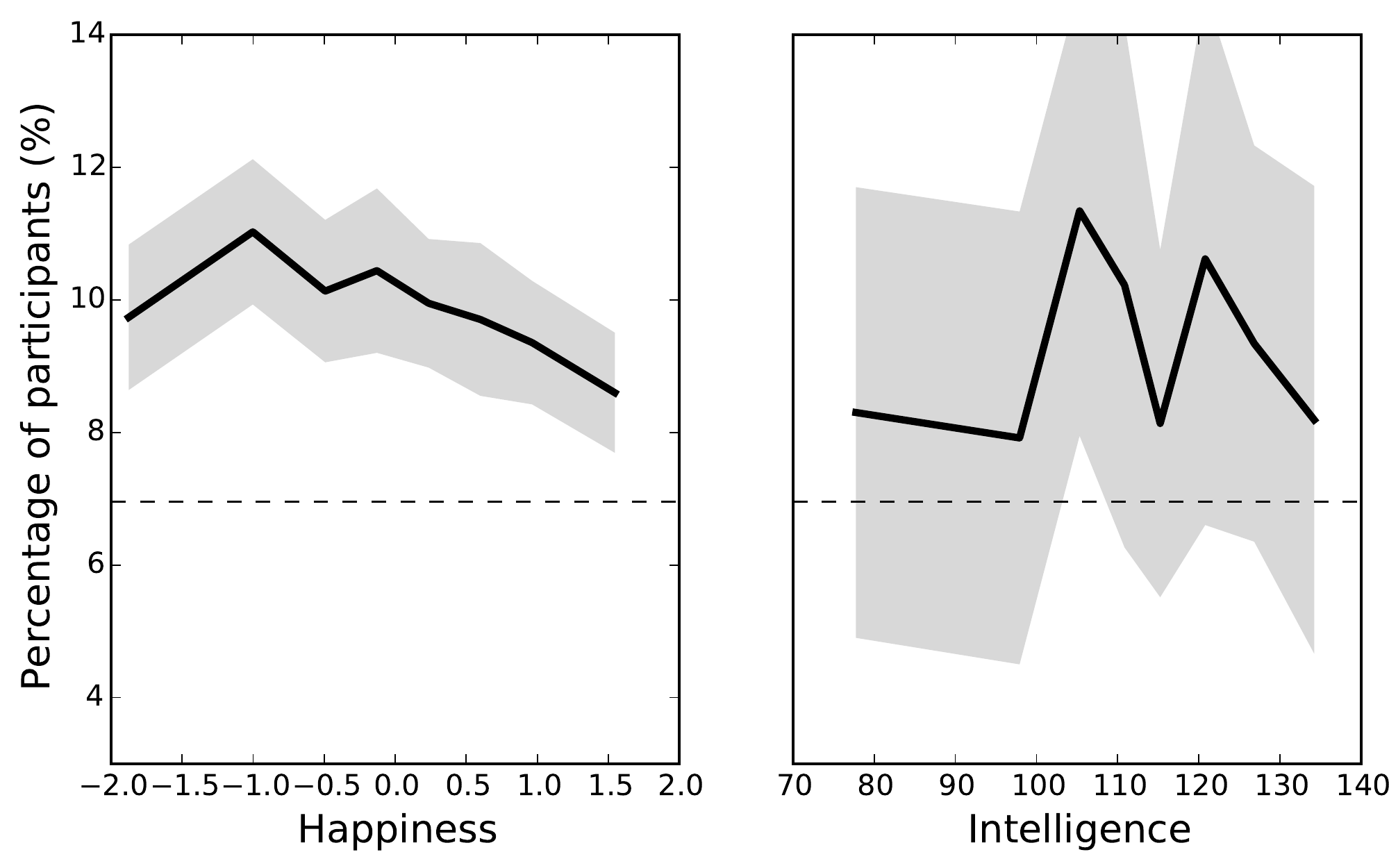}
\caption{Relationship between the Rainbow campaign participation, happiness and intelligence. Dashed line represents the overall percentage of campaign participants in our sample (6.96\%). Grey ribbon represents the 95\% confidence interval.}
\label{fig:rb_swliq}
\end{figure}

%% file: 040personality.tex
Studies have shown that personality highly correlates with a broad range of beliefs and attitudes, including those related to religion \cite{argyle1975social} and political views \cite{caprara2006personality,saucier2000isms}. Here, we analyze the relationship between the Rainbow campaign participation and personality traits such as openness, conscientiousness, extraversion, agreeableness, and neuroticism. 

\begin{figure*}
\centering
\includegraphics[width=0.99\linewidth]{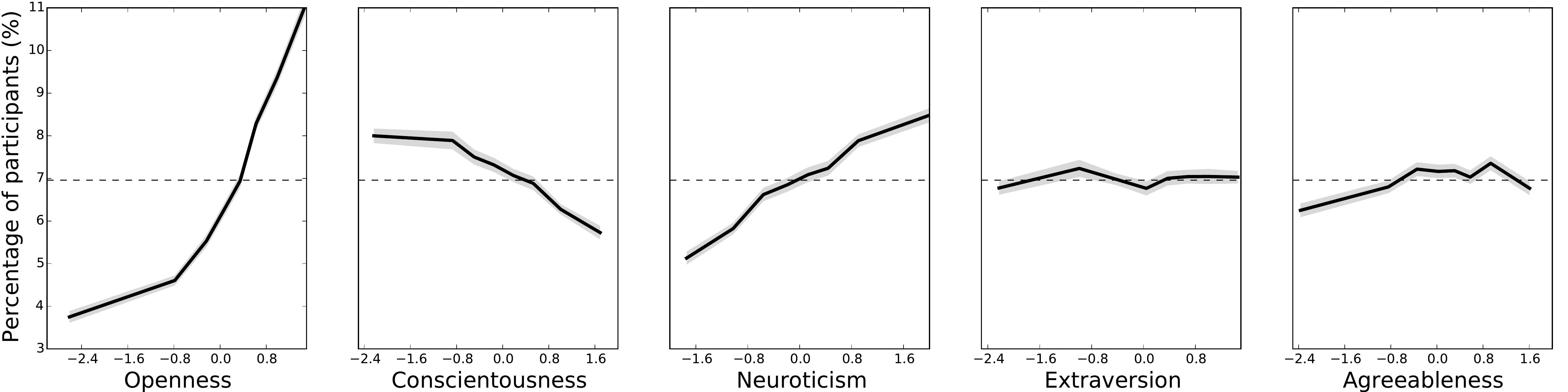}
\caption{Relationship between Rainbow campaign participation and Five Factor Model of personality traits. Grey ribbons represent 95\% confidence interval.}
\vspace{0.1in}
\label{fig:rb_ffm}
\end{figure*}

Figure~\ref{fig:rb_ffm} shows the relationship between campaign participation and the five personality traits. Scores on each trait are binned into eight quantiles. Openness is most strongly correlated with the campaign participation: open-minded (i.e., liberal and unconventional) individuals were almost three times more likely to join the campaign (11.06\%) compared to their conservative counterparts (3.75\%). This insight is similar to previous findings which demonstrated that openness is a very strong predictor of political views~\cite{saucier2000isms}. The campaign participation was negatively correlated with Conscientiousness. People with low conscientiousness (i.e., spontaneous and impulsive individuals) showed higher rates of participation than those ranked high on this trait (i.e., self-disciplined and dutiful). Finally, campaign participants are more likely to score high on neuroticism (i.e., tendency to experience unpleasant emotions, such as anger, anxiety, depression, and vulnerability). Neurotic people were nearly two times more likely (8.48\% versus 5.13\%) to support the campaign than more emotionally stable individuals. This result indicates that emotional individuals might be more likely to take public stances on controversial issues. There were no strong correlations between the two remaining personality traits, extraversion and agreeableness, and campaign participation. 

\subsection{Happiness and Intelligence}
\label{sec:happy}

We observed no significant relationship between campaign participation and levels of happiness and intelligence (Figure \ref{fig:rb_swliq}). Interestingly, however, campaign participation was positively correlated with the mere fact of taking SWLS questionnaire or Intelligence test 
Given that myPersonality application on Facebook required its users to take the personality test before they could access other measures, this indicates that those users who were motivated to take additional tests, were also more likely to participate in the campaign.

%% file: 050social.tex
In this section, we analyze the social ties among campaign participants and study the homogeneity of their egocentric social networks. 
\vspace{-0.15in}
\paragraph{Social Ties}
Studies have shown that behaviors, attitudes and ideologies spread through social connections~\cite{bakshy2015exposure,de2013anatomy} and that people tend to interact with others who are similar to them~\cite{anderson2015global,mcpherson2001birds}. 
The same phenomenon applies to the support for same-sex marriage. This was shown by
\citeauthor{adamic2015diffusion}, \citeyear{adamic2015diffusion} who studied a phenomenon similar to the Facebook Rainbow campaign: a 2014 viral diffusion of the equal-sign-themed profile picture in support of the same-sex marriage equality. Their results show that people were more likely to adopt the equal-sign icon if they observed their friends doing so. 

Our research supports the findings of the aforementioned study. Our data shows that 12.39\% (95\% CI\footnote{CI stands for Confidence Interval.} [12.32, 12.76])
of friends of campaign participants were also campaign participants. Non-participants only had 8.31\% (95\% CI [8.18, 8.44])
of friends who participated in the campaign.\footnote{Both percentages are higher than the overall percentage of campaign participants in our dataset (6.96\%). This is because the users for whom we had network data were more likely to support the Rainbow campaign.} 

We further explored the relationship between the likelihood of campaign participation and the percentage of his/her friends who were campaign participants. We included only those users for whom we had at least 10 friendship links available in this analysis. Figure \ref{fig:perOfFF} shows that the likelihood of joining the campaign grows linearly with the percentage of friends who themselves are participants. 

\vspace{-0.15in}
\paragraph{Homogeneity in Social Ties}
\label{sec:social_propensity}
We also probed the homogeneity of campaign participants' networks. First, we grouped users based on their religion, political view, and gender. Next, we measured the fraction of people who belonged to the same category among their friends. 

Results presented in Figure~\ref{fig:diff_pie} show that male campaign participants exhibited a higher degree of homogeneity (48.02\% of their friends were also male) compared to males who were non-participants (42.68\% of their friends were also male), resulting in about 5\% difference. No such difference in network homogeneity was observed for females.

On the contrary, networks of conservative and christian campaign participants were more heterogeneous. Conservative campaign participants had $12.21\%$ more liberal friends than conservative non-participants. This implies that conservatives surrounded by many liberals are, in fact, closer to the liberal point of view. Similarly, although social ties of Christians were predominantly homogeneous ($87.67\%$ of their friends were Christians), Christian campaign participants had $4.68\%$ more friends who were non-believers, meaning that their networks were more heterogeneous and they were closer to the non-believers community.  
No such differences in homogeneity were found between participant and non-participant populations of democrats and non-believers. 

%% file: 060like.tex
This section explores the relationship between campaign participation and Facebook Likes, which serve as a convenient proxy for users' interests, preferences, opinions, and affiliations. Here, we compute the Phi coefficient to test the strength of relationship between user Likes and the campaign participation. Phi coefficient is a measure of association between two binary variables~\cite{cramer1999mathematical}. Figure \ref{fig:like_dis} shows the distribution of the Phi coefficient in our sample. The majority of Likes have a Phi coefficient value $\approx 0$, which means that there is hardly any correlation between users liking pages and campaign participation. However, there are still a fair amount of Likes which are related to campaign participation.

\begin{figure}
\centering
\includegraphics[width=0.99\linewidth]{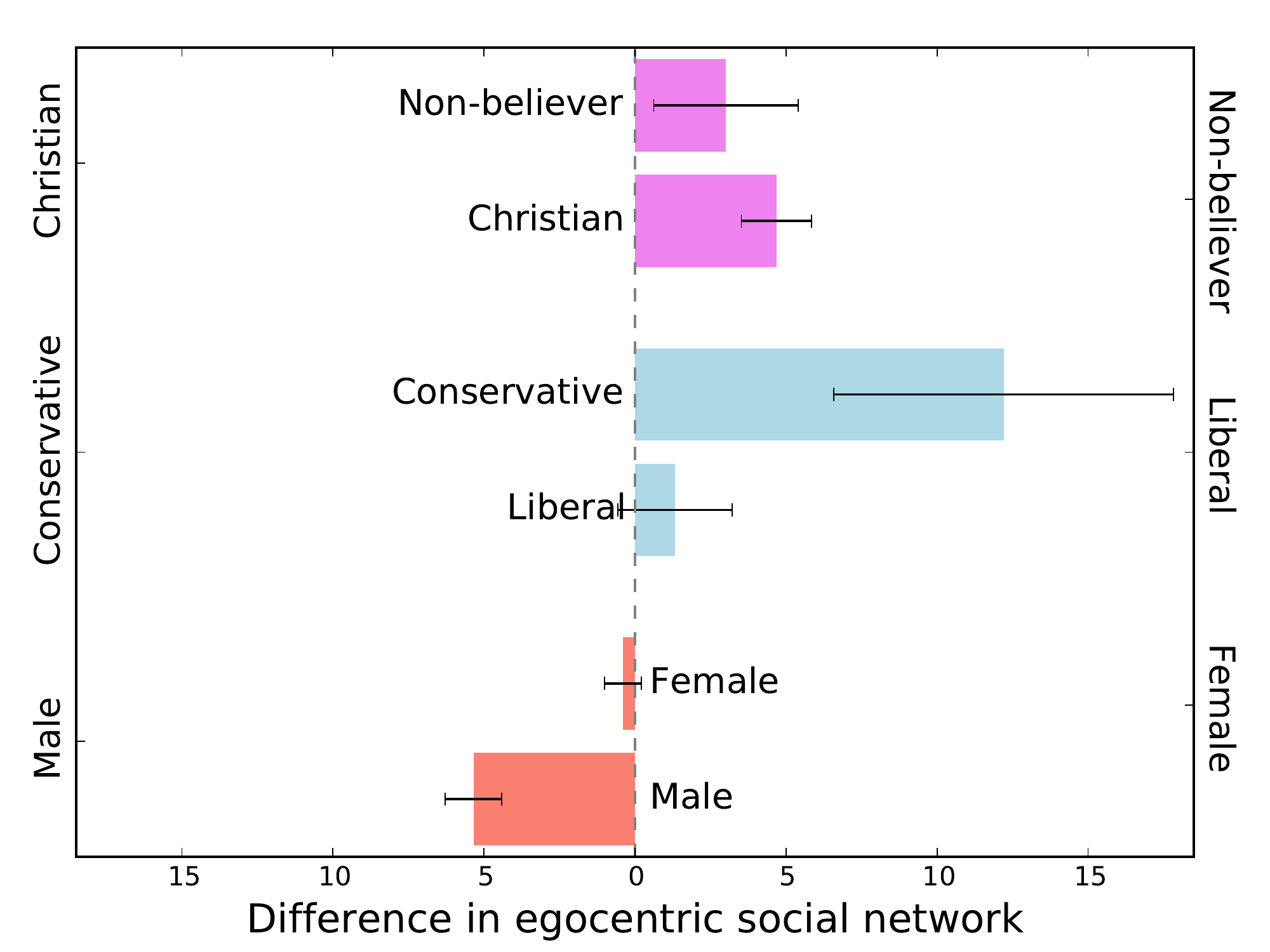}
\caption{Differences in network homogeneity between male, conservative and Christian participants and the corresponding non-participants. For example, male participants of the campaign have 5.34\% more male friends than non-participants. Error bars represent 95\% confidence intervals of differences between two percentages.}
\label{fig:diff_pie}
\vspace{-0.15in}
\end{figure}


Table~\ref{tab:like} shows the names of top 20 Likes which are highly correlated (both positively and negatively) with campaign participation. Some interesting patterns emerge. Likes that are positively associated with campaign participation include those directly related to LGBT rights (e.g., "Gay Marriage", "Human Rights Campaign", "NO H8 Campaign", "Let Constance Take Her Girlfriend to Prom!", "The Gay, Lesbian \& Straight Education Network (GLSEN)", "Gay \& Lesbian Victory Fund", "Repeal Don't Ask Don't Tell"), and those related to openly gay-friendly artists, such as Lady Gaga, Kathy Griffin, Ellen DeGeneres; and TV shows, such as "Glee" (includes many LGBT characters) and "The L Word" (a story of lesbians and bisexuals). Other associations were somewhat less obvious. For example, books such as "Harry Potter" and "Alice in Wonderland", which do not seem to be directly related to LGBT rights, were positively correlated with the campaign participation. They may, to some extent, appeal to liberal and open-minded (and thus more LGBT friendly) audiences.

Among Likes that are most negatively correlated with campaign participation, meaning strongly associated with non-participation, there is an abundance of religious and conservative ("The Bible", "Being Conservative", "Jesus Daily", "I Can Do All Things Through Christ Who Strengthens Me") entities which is consistent with findings discussed in previous sections. There are also a large number of Likes related to African American artists and sports stars such as Wiz Khalifa, Kevin Hart, Michael Jordan, Lebron James, Lil Wayne, and Young Jeeze. Some of them (e.g., Kobe Bryant, Trey Songz) were suspected of being gay but later claimed to be straight\footnote{\tiny{http://hollywoodlife.com/2014/03/26/trey-songz-gay-tweet-prank-rumors/}
} and few others were once involved in anti-gay issues. \footnote{\tiny{http://articles.latimes.com/print/2011/apr/13/sports/la-sp-kobe-bryant-lakers-20110414}}
\begin{figure}
\centering
\includegraphics[width=0.99\linewidth]{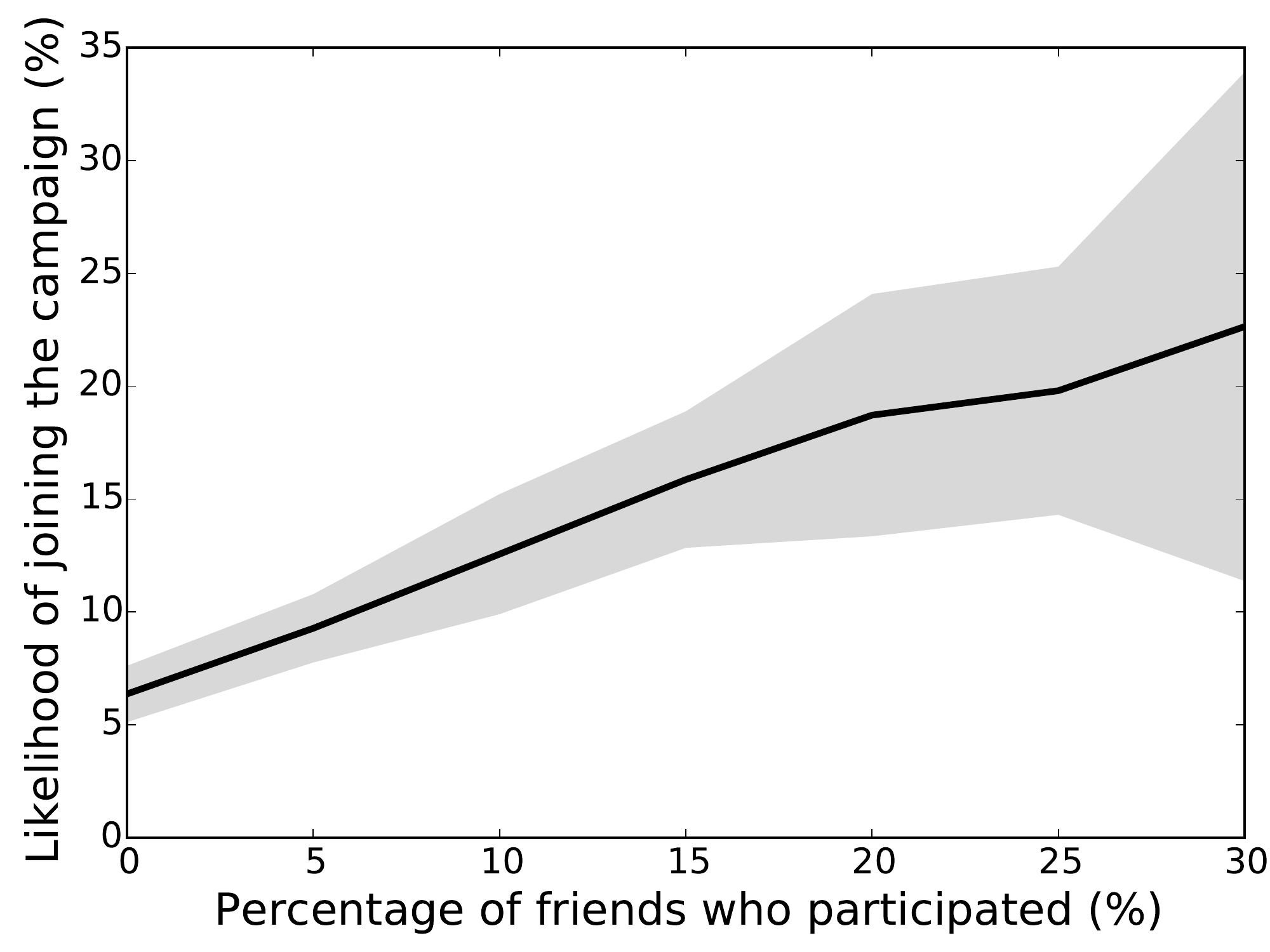}
\caption{
Percentage of friends who participated in the campaign versus the likelihood of joining the campaign. The grey ribbon area represents the 95\% confidence interval.}
\label{fig:perOfFF}
\end{figure}

%% file: 070prediction.tex
In this section, we discuss our models aimed at predicting participation in the Facebook Rainbow campaign based on the features discussed in the previous sections.
Results presented here were obtained using a Logistic Regression model with 10-fold cross-validation; other machine learning models, such as Random Forest and SVM, produced similar results. Since the campaign participants represent a minority of users (6.96\%), the prediction performance is reported in terms of the AUC.

\begin{figure}
\centering
\includegraphics[width=0.99\linewidth]{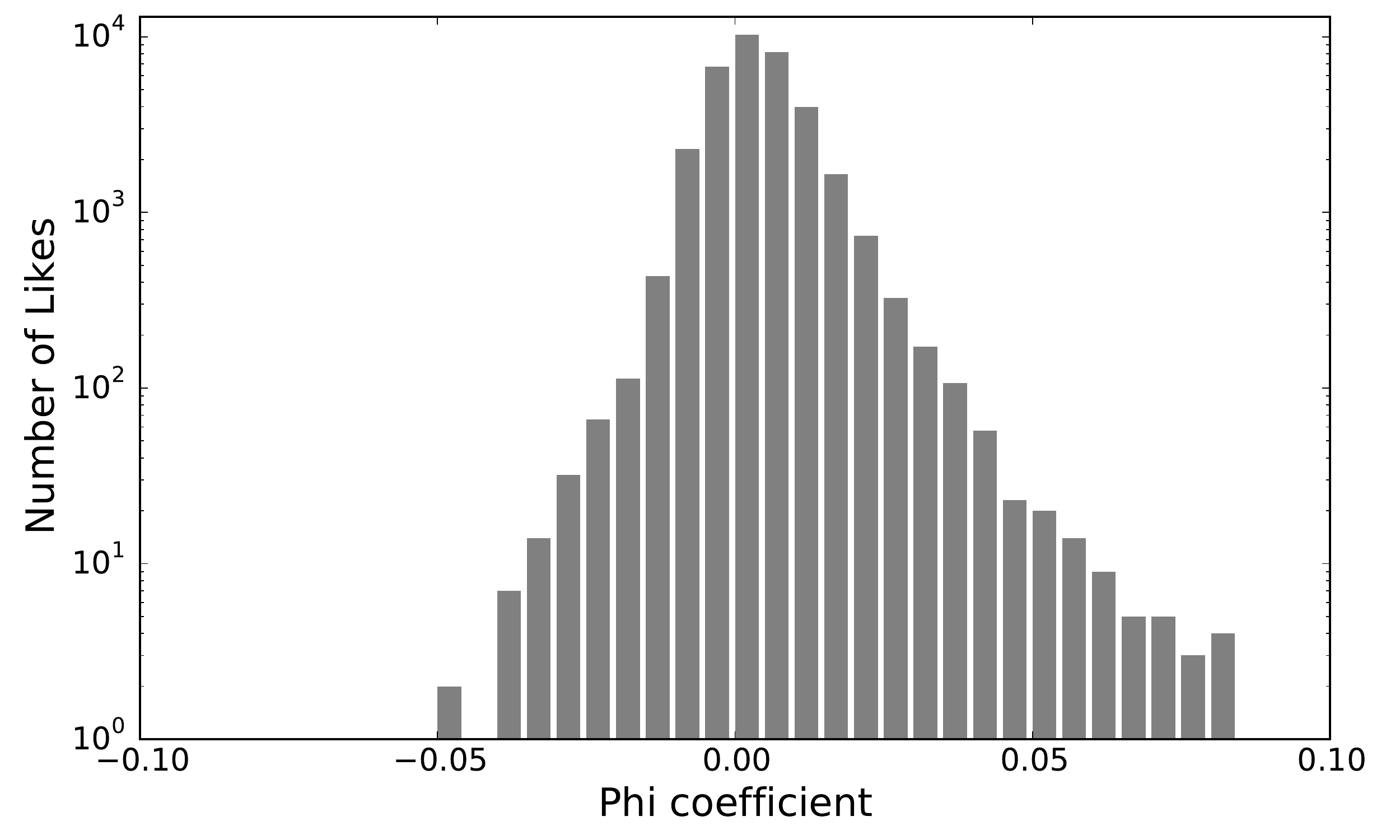}
\caption{Distribution of the Phi coefficient describing the strength of association between Facebook Likes and campaign participation. Note that the number of Likes is on a logarithmic scale.}
\label{fig:like_dis}
\end{figure}

\subsection{Features \& Pre-processing}
We use the following features discussed in the previous sections: five personality traits, demographic attributes (age,\footnote{We use 8 quantile bins to discretize the age variable as its relationship with the likelihood of campaign participation is non-linear (see Figure \ref{fig:rb_age}).} gender, and state), views (political and religious), percentage of friends who participated, happiness, intelligence, and Facebook Likes. Specifically, as we store Facebook Like information using high dimensional vectors, they are preprocessed following the procedure described in~\cite{kosinski2013private}. First, we construct a user-Like matrix, where the columns represent Likes and the rows represent users. An entry in this matrix $(u,l)$ is set to 1 if user $u$ expressed interest in Like $l$, otherwise it is set to 0. The dimensionality of the user-Like matrix was reduced using Singular Value Decomposition. The resulting $k=50$ dimensions were used as features in the prediction tasks. 

\begin{table}
\centering
\caption{Accuracy of predicting campaign participation in terms of AUC}
\vspace{0.1in}
\begin{tabular}{lc} \specialrule{1.5pt}{0pt}{0pt}
\multicolumn{2}{c}{\textbf{Individual Features \& Feature Sets}}\\ \hline
\textbf{Feature}&\textbf{AUC}\\ \hline
{Likes} & {0.73} \\ \hline
{Views (all)} & {0.70} \\ 
\textit{Political View }& 0.64\\ 
\textit{Religion} &0.62\\ \hline
{\% of participating friends } & {0.62} \\ \hline
{Personality (all)} & {0.61} \\ 
\textit{Openness} & {0.59} \\ 
\textit{Conscientiousness} & {0.54} \\ 
\textit{Extraversion} & {0.51} \\ 
\textit{Agreeablnees} & {0.51} \\ 
\textit{Neuroticism} & {0.55} \\ \hline
{Demographics (all)} & {0.59} \\ 
\textit{Gender} & 0.56\\
\textit{State} & 0.55\\
\textit{Age} & 0.53\\ \hline
Happiness & 0.51\\ \hline
Intelligence & 0.50\\ \specialrule{1.5pt}{0pt}{0pt}

\multicolumn{2}{c}{\textbf{Combined Features}}\\ \hline
\textbf{Feature}&\textbf{AUC}\\ \hline
Likes + Views + Personality &0.76\\
Likes + Views  &0.75\\ 
Views + Personality & 0.71\\ \specialrule{1.5pt}{0pt}{0pt} 
\end{tabular}
\label{tab:predictions}
\end{table}

\subsection{Prediction Accuracy}
Prediction accuracies achieved using both individual features and combined feature sets are presented in Table~\ref{tab:predictions}. The highest accuracy among individual features (AUC=0.73) was achieved for SVD dimensions ($k=50$) representing user Facebook Likes. This is a remarkable performance given that those Likes were recorded between 2009 and 2011 before the start of the Rainbow campaign. 
This result nicely illustrates the potential of predictions based on digital footprints in general and Likes in particular (See also: \citeauthor{kosinski2013private} \citeyear{kosinski2013private}, \citeauthor{youyou2015computer} \citeyear{youyou2015computer}).

Political views (AUC=0.64) and religion (AUC=0.62) were also reasonably predictive of the campaign participation. Combining those variables boosts the AUC to 0.70, suggesting that religion and political views are, to a large extent, independently predicting campaign participation. 
The next most predictive feature (AUC=0.62) was the fraction of friends who were campaign participants (Refer to the Section on Social Networks for more details). Combined personality traits resulted in an AUC=0.61. Individual personality traits, on the other hand, exhibited slightly lower predictive power. It can be seen from Table \ref{tab:predictions} that among all the individual personality traits, openness is the best indicator of campaign participation, followed by conscientiousness and neuroticism.

\begin{table*}
\centering
\caption{Facebook Likes most positively and negatively associated with Rainbow campaign participation}
\vspace{0.1in}
\begin{tabular}{ll} \hline
20 Positively Correlated Likes&20 Negatively Correlated Likes\\ \hline 
Human Rights Campaign & The Bible\\ 
NO H8 Campaign &Being Conservative\\ 
Girls Who Like Boys Who Like Boys &Lebron James\\ 
Gay Marriage &Wiz Khalifa\\ 
Let Constance Take Her Girlfriend to Prom! &Bible\\ 
Day of Silence &Don't give up on God because he never gave up on you..:) \\ 
Ellen DeGeneres &ESPN\\ 
Lady Gaga &Kevin Hart\\ 
Glee &Michael Jordan\\ 
Harry Potter &Trey Songz\\
The Gay, Lesbian \& Straight Education Network (GLSEN) & Jesus Daily \\
The Ellen DeGeneres Show & Conservative\\
Repeal Don't Ask Don't Tell & "I'm Proud To Be Christian" by Aaron Chavez\\
Alice in Wonderland & Waka Flocka Flame  \\
Kathy Griffin & Kobe Bryant\\
Hot Topic & Gucci Mane \\
FCKH8.com & Nike Football\\
The L Word & Lil Wayne\\
American Foundation for Equal Rights (AFER) & Young Jeezy\\
Gay \& Lesbian Victory Fund & Hunting\\
No on Prop 8 | Don't Eliminate Marriage for Anyone & I Can Do All Things Through Christ Who Strengthens Me\\ 
\hline
\end{tabular}
\label{tab:like}
\end{table*}

Gender (AUC=0.56), state  (AUC=0.55), age (AUC=0.53) identified campaign participants with moderate success consistent with the results presented in the Section on demographics. Combining these three demographic features gave a predictive accuracy of AUC$\approx$0.60, which is comparable to that of combined personality traits and percentage of participating friends. As previously discussed in the Section on Happiness and Intelligence, the performance of the models trained using happiness (AUC=0.51) and intelligence (AUC=0.50) turned out to exhibit a similar performance as that of a random baseline.

In practice, features belonging to separate categories are often combined to maximize the model performance. The benefits of combining multiple features can be clearly seen in the results included at the bottom of Table~\ref{tab:predictions}. The model combining Likes, views, and personality turns out to be very predictive of the participation (AUC$=0.76$). Removing personality from the feature set does not significantly decrease the predictive power (AUC=0.75); removing Likes, instead, leads to a higher decline (AUC=0.71). 
This confirms previous findings~\cite{kosinski2013private,youyou2015computer} showing that the information contained in personality scores is also encoded in user Facebook Likes.
\footnote{We restricted our analysis to only three feature combinations because combining multiple features limits the number of available cases, due to missing values.}

%% file: 090conclusion.tex
Pictivism, a new way of engaging in \emph{activism} by changing online profile \emph{pictures}, has gained enormous traction in the recent past. Several campaigns which fall under the umbrella of pictivism are being widely adopted by online users all over the world primarily because they require minimal effort on the part of individuals. Understanding the characteristics of the participants of such campaigns is crucial to the sustainment of online social activism. On the other hand, marriage equality is one of the most important and divisive social issues of our times. Improving our understanding of people's attitudes towards this issue is beneficial both to the LGBT community and society as a whole. This work bridges the gap between two such very contemporary and diverse research directions. 

In this paper, we analyze one of the most massive instances of pictivism: the Facebook Rainbow campaign. This work explores the relationship between participation in the Rainbow campaign and various attributes such as psycho-demographic traits, network ties, and personal preferences. Our analysis of the data unearthed various interesting insights. We showed that non-believers, democrats, women and those between 30 and 40 years of age were more likely to participate in the campaign compared to their counterparts. We also found that participation rates were higher among the Civil War era Union states and those states which legalized same-sex marriage prior to the Supreme Court verdict. Our analysis also revealed that personality traits such as openness and neuroticism are positively correlated with the likelihood of participation, while conscientiousness was negatively correlated. Further, the participation rates were positively correlated with the percentage of friends participating in the campaign. Lastly, we also found that Christian and republican campaign participants had less homogeneous social networks (i.e., they were surrounded by more non-believers and democrats than their non-participating counterparts).

We also developed a logistic regression model to predict the campaign participation of individual users. We found that Facebook Likes (used to express users' interests and preferences) are highly predictive of campaign participation (AUC = 0.73), followed by the fraction of participating friends (AUC = 0.62), personality profiles (AUC = 0.61), and demographic traits (AUC$\approx$0.60). Combining multiple features (Likes, Views, and Personality) allowed the model to reach a relatively high prediction accuracy (AUC = 0.76).  

Participation in the Rainbow campaign is a proxy not only for an individual's support towards same-sex marriages, but also an individual's willingness to take an active stance in an important public debate. Consequently, in addition to facilitating social activism, social media provides researchers with a medium to study social activism on a scale inconceivable just a decade ago. While extensive theoretical and empirical research on social activism has been undertaken in the fields of political science and sociology, online campaigns provide us with a rich and low-cost opportunity to study social activism through the lens of large-scale data-driven analysis. We hope that this work encourages researchers to use social media as a valuable tool to understand social campaigns. 
\vspace{-0.15in}
\paragraph{Future Work} 
This work raises several interesting questions some of which could not be addressed here partly due to limitations of the available data. Firstly, it would be interesting to analyze the diffusion of the campaign participation. (See \citeauthor{adamic2015diffusion}, \citeyear{adamic2015diffusion} for an example of such study.) Due to limited access to network and time-stamp data, diffusion could not be studied in this work. Second, similar studies aimed at other social campaigns could help identify psycho-demographic correlates of social activism in general.